\documentclass[doublecol,citesort]{epl2} 
\usepackage{diagbox}
\usepackage{caption}
\usepackage{subcaption}
\usepackage[bookmarks,colorlinks]{hyperref}
\newcommand\scalemath[2]{\scalebox{#1}{\mbox{\ensuremath{\displaystyle #2}}}}
\usepackage[normalem]{ulem}

\title{Accelerations of large inertial particles in turbulence
}
\shorttitle{Large particles in turbulence} 

\author{Yaning Fan\inst{1},
Cheng Wang\inst{1},
Linfeng Jiang\inst{1,2},
Chao Sun\inst{1}\footnote{chaosun@tsinghua.edu.cn},
Enrico  Calzavarini\inst{3}\footnote{enrico.calzavarini@univ-lille.fr}}
\shortauthor{Y. Fan  \etal}

\institute{  
\inst{1}New Cornerstone Science Laboratory, Center for Combustion Energy, Key Laboratory for Thermal Science and Power Engineering of MoE, Dept. of Energy and Power Engineering, Tsinghua University, 100084 Beijing, China\\
  \inst{2} Physics of Fluids Group, University of Twente, 7500 AE Enschede, The Netherlands\\  
  \inst{3} Universit$\acute{e}$ de Lille, Unit$\acute{e}$ de M$\acute{e}$canique de Lille - J. Boussinesq (UML) ULR 7512, F 59000 Lille, France\\
}

\abstract{Understanding the dynamics of material objects advected by turbulent flows is a long standing question in fluid dynamics.
In this perspective article we focus on the  characterization of the statistical properties of non-interacting finite-sized massive spherical particles advected by a vigorous turbulent flow.
We study the fluctuations and temporal correlations of particle accelerations and explore their behaviours with respect to the particle size and the particle mass density by means of fully-resolved numerical simulations. 
We observe that the measured trends can not be interpreted as the simple multiplicative combination of the two dominant effects: the spatial filtering of fluid accelerations and the added-mass-adjusted fluid-to-particle density ratio. We argue that other hydrodynamical forces or effects, e.g. preferential flow sampling, have still a significant role even at the largest particle sizes, which rich here the integral scale of turbulence.}

\begin{document}

\maketitle

\section{Introduction}
Fluid dynamics turbulence is a lasting challenge in science. Rather than representing a single fundamental question  - the problem's definition itself changed over epochs and disciplines
\cite{Eckert_book} -  it is a faceted topic with ramifications into a plethora of open issues and applications. 
Among the many problems connected to turbulent flows a long standing one concerns the description of the transport by the flow of material objects\cite{BrandtCollettiARFM22}. 
Clearly, the study of the forces exerted on a body immersed in a flow is a classic topic in the broader field of fluid dynamics, and many researchers have devoted their attention to it. 
When considering the case of spherical bodies, we can refer to Stokes's work on drag force in creeping flows, followed by  studies by Oseen, Boussinesq, and Basset on inertial and unsteady corrections (history force) to drag. Tchen, Corrsin, and Lumley investigated the pressure gradient force, while Auton focused on added-mass and lift forces in the context of rotational potential flows (see \cite{Michaelides97} for an historical overview). The unified modern formulation on the dynamics of a spherical particle in an unsteady and non uniform viscous flow is due to Maxey \& Riley and Gatignol (MRG equation) \cite{maxey-riley-1983,gatignol-1983}\footnote{The same equation, without Fax\'en corrections \cite{faxen1924resistance}, was derived by Shu-Tang Tsai in 1957 and published in Chinese language, see \cite{Tsai2022} for its recent translation.}. However, the complexity of the problem increases considerably when the carrying flow is turbulent, due to its non-smooth and erratic characteristics both in time and space. In this case, any kind of quantitative investigation must embrace the statistical approach \cite{ToschiBoden2009,BecetalARFM2024}.

In the last decades, a large body of studies has been dedicated to the problem of particles advected by turbulence. This is largely attributed to the emerging methods of experimental fluid/particle tracking by digital cameras \cite{adrian1991particle,2001NatureParticle,voth_laporta_crawford_alexander_bodenschatz_2002} and by the ever increasing high-performance numerical simulations \cite{ToschiBoden2009}. While the phenomenology is presently relatively well explored for particles whose size is of the order of the dissipative scale and the associated particle Reynolds number is small, there are fewer studies on less idealized objects of larger size \cite{BURTON_EATON_2005, QureshiPRL2007,xu2008motion, Naso2010, LUCCI_FERRANTE_ELGHOBASHI_2010, volk2011dynamics,MathaiPRF2015,uhlmann_chouippe_2017} or with non regular shape \cite{VothSoldatiARFM2017,verhille_2022} or inhomogeneous mass density \cite{will_krug_2021}.
The question is important for applications, among many we wish to mention the topical issue of the dispersion in turbulent ocean of plastic debris \cite{LawARMS2017}. Such debris encompass ranges of scales from the dissipative to the inertial ones, have a variety of shapes and buoyant properties \cite{SutherlandPRF2023} and can undergo turbulence induced fragmentation \cite{PhysRevFluids.6.024601}. Advances in the comprehension of the problem are crucially linked to a better insight into their translational dynamics. 

In this article we aim at performing a step forward in the understanding of this complex phenomenon. We aim at highlighting how the particle mass density affects the acceleration properties for particles whose size is larger than the dissipative scale of turbulence. Our approach is computational, and makes use of numerical simulations that are capable to resolve all the spatio-temporal scales active in the problem. Due to the wide range of scales at work, these simulations are possible at the price of considering only single (or at best few non-interacting) particles and moderate turbulent flow intensities. Despite this limitation, our study allows to identify trends that are expected to be valid also at larger turbulent flow Reynolds numbers and it provides new benchmark data for future studies.

\section{Theoretical considerations}
Recent experimental and numerical studies have shown that the translational \cite{Brown2009prl,volk2011dynamics}, and in part the rotational \cite{jiang_wang_liu_sun_calzavarini_2022}, statistical properties of inertial-scale-sized neutrally buoyant particles advected by a turbulent flow can be explained in terms of a \textit{coarse-graining} effect of the underlying turbulent flow. In other words the particle acceleration behaves - statistically - the same as the spatially-filtered fluid acceleration field unperturbed by the particle.
This mechanism corresponds to assuming that the particle feedback on the carrying turbulent flow is negligible. 
A neutral particle has on average a small slip velocity compared to the surrounding fluid \cite{cisse2013slipping}, it is therefore reasonable to assume that the dissipative (surface) force associated to viscous drag is sub-leading with respect to the inertial (volumetric) one.
In summary it seems reasonable to state that the acceleration of a particle of typical size $d$  goes as  $\bm{a}_d \sim \langle D_t\bm{u} \rangle_{V}$ where $D_t \bm{u} = \partial_t \bm{u} + \bm{u} \cdot \nabla \bm{u}$ is the fluid acceleration field and $\langle \ldots \rangle_{V}$ denotes a spatial average over a volume ($V$) equivalent to the one of the particle.

When the particle is non-neutrally buoyant this scenario is complicated by two factors, on one hand the different inertia between the particle and the fluid tends to suppress/enhance (respectively for heavy/light particles) the fluid accelerations of the surrounding carrying flow. This effect is not plainly proportional to fluid-to-particle density ratio, $\rho_f/\rho_p$, but is instead proportional to the parameter $\beta = 3 \rho_f/(\rho_f + 2 \rho_p)$ because of the added-mass force exerted by the fluid on the particle \cite{autonJFM1988}. Note that $\beta$ varies in the interval $\left[0,3\right]$, and the limits  correspond to the cases of very massive particles (ballistic limit) and of very light particles where the inertia is all in the displaced surrounding fluid (such as for the case of air bubbles in water\cite{vargheseAR2020}); $\beta=1$ identifies the neutrally buoyant case.
The second factor is the occurrence of Archimedes buoyancy, which leads to an extra acceleration term of the form $(1-\beta)\bf{g}$. 
In this study for simplicity we neglect the effect of gravity, and this is always possible for sufficiently intense turbulent flows\footnote{However, note that at increasing the particle size in a fixed intensity turbulent flow, the relative importance of the coarse grained fluid acceleration decreases with respect to the  buoyancy.}.
Adding together the mentioned volume and surface forces the following model as been put forward for the motion of finite-sized particles \cite{calzavarini2009acceleration,calzavarini2012impact}: 
\begin{equation}
\scalemath{0.92}{
\ddot{\bm{x}} = \beta \left( \langle D_t \bm{u}\rangle_V + \frac{12\ \nu\ c(Re_d)}{d^2}(\langle \bm{u} \rangle_S-\dot{\bm{x}}) \right) + (1-\beta)\bm{g}.
}
\label{eq:faxen}
\end{equation}
This is an adaptation of the  MRG equation, which retains the so called Fax\'en terms \cite{gatignol-1983}, with the addition of the empirical Shiller-Naumann drag correction $c(Re_d) = 1 + 0.15 Re_d^{0.687}$ where $Re_d=||\langle \bm{u} \rangle_S - \dot{\bm{x}}|| d/ \nu$ is the instantaneous particle Reynolds number and the omission of the history force \textcolor{black}{(justified for non fastly settling particles \cite{vanHinsberg2017})}. 
The symbols $\langle \ldots \rangle_S$ denotes, similarly to the volume mean,  a spatial average over a surface (S) equivalent to the one of the particle. These averages quantifiy the effect of the local flow non uniformity at the scale of the particle. If the particle is small they account simply for the effect of the curvature (i.e. the laplacian) of the local flow velocity and acceleration fields, if the particle is large they include higher even-order spatial-derivatives, see also \cite{BhargavPRF2021}. This model, known as Fax\'en corrected (FC) model  \cite{calzavarini2009acceleration,calzavarini2012impact}, predicts qualitatively a series of trends in the  single- and two-time acceleration statistics, that will be discussed later on in this article. 

We can now ask, what are the statistical features of the particle acceleration in a turbulent flow for the non-neutral ($\beta \neq 1$) case? 
When particles are below the dissipative scale \textcolor{black}{and the flow accelerations are large compared to gravity}, the Fax\'en and the Shiller-Naumann corrections can be neglected in eq.~(\ref{eq:faxen}) and 
expanding in the small parameter $d^2/(12 \nu \beta) \ll1$(the drag response time) one obtains:
\begin{equation}
\ddot{\bm{x}} \simeq D_t\bm{u}  +  \frac{d^2}{12 \nu \beta}(\beta-1)  \left( D_t\bm{u} \cdot \nabla \bm{u} + D_t^2\bm{u} \right).
\label{eq:acc-small}
\end{equation}
This tells that the particle acceleration variance begins to deviate from the one of a fluid tracer quadratically with its size $d$ and only when $\beta \neq 1$. 
For finite-sized particles a similar perturbative estimate is not possible. However, the drag response time becomes long and one can assume that the associated force becomes negligible.
All these considerations lead us to guess that inertia - the term $\beta \langle D_t \bm{u}\rangle_V$ in (\ref{eq:faxen}) - is the leading effect in determining the statistics of the acceleration of large particles. 
This study aims at testing this hypothesis and discussing its implications. 

\section{Methods: Fully resolved numerical study}
We address the above questions by solving the fluid-particle coupled problem which comprises the incompressible Navier-Stokes equations for the fluid dynamics and the Newton-Euler equation for the particle motion, with the addition of no-slip boundary conditions at their interfaces.
 The numerical methods, based on the coupled Lattice-Boltzmann and Immersed-Boundary algorithms, have been already described elsewhere \cite{Calzavarini_SI2019,jiang_wang_liu_sun_calzavarini_2022}. For the present study, we carried on new validations on the dynamics of non neutrally buoyant particles. Besides resolution convergence checks we verified  that in the presence of gravity the particle trajectory in still fluid agrees with the known settling dynamics. 
The spatial domain where the turbulent flow takes place is a tri-periodic cube.
The Taylor-scale based Reynolds number is kept at  $Re_{\lambda}\simeq 32$ by a \textcolor{black}{constant-power large-scale forcing (same as in \cite{PerlekarPF2012})}. The values of  the relevant numerical/ physical turbulent flow scales are reported in Table \ref{table:NS}.
	\begin{table}
		\begin{center}
		\resizebox{\columnwidth}{!}{
			\begin{tabular}{ccccccc}
				$N^3$&$\eta/\Delta x$& $\tau_\eta/\Delta t$ & $L/\eta$ & $T_L/\tau_\eta$ & $\lambda/\eta$ & $Re_\lambda$\\
				\hline
				$128^3 (256^3)$ &  $2.1 (4.2)$ & $153 (607)$ & $24.1$ & $12.4$ & $11.3$ &  $32$
			\end{tabular}
			}
			\caption{Parameter and relevant scales of the simulated turbulent flow. $N^3$: number of spatial grid points (the larger resolution is used for $\beta>1$ particles); $\eta$: Kolmogorov dissipation length scale in grid space units $\Delta x$; $\tau_\eta$: Kolmogorov time scale in time-step units $\Delta t$, $L$: integral scale; $T_L$: large-eddy turnover time; $\lambda$: Taylor micro-scale; $Re_\lambda$: Taylor-Reynolds number.}
			\label{table:NS}
			
		\end{center}
	\end{table}
The control particle parameters are varied in the range $d/\eta=[ 6.5,18.7]$, which comprises the inertial range of the considered turbulent flow, and $\rho_p/\rho_f = [0.4,10]$ (corresponding to $\beta=[0.14, 1.67]$).  The particle are evolved for $O(10^2)$ large-eddy turnover times. Each simulation contains just one particle, but multiple independent runs are performed for each particle case \textcolor{black}{in order to improve the statistical convergence of the analysis.}


\section{Results on acceleration statistics}
\subsection{Acceleration variance}
The first quantity we focus on is the single-component acceleration variance, $\langle a_{d,i}^2 \rangle$ where $\langle \ldots \rangle$ denotes here the time and ensemble average over independent particle trajectories (all the measurements are also given in Table \ref{table:acce_variance}).
	\begin{table}
		\begin{center}
        \resizebox{\columnwidth}{!}{
            \begin{tabular}{|c|cccccc|}
            \hline
                \diagbox{$\beta$}{$d/\eta$} & 6.5419 & 9.3444 & 11.2119 & 13.0785 & 15.8774 & 18.6739 \\ \hline
                1.6667 & 0.4538 & 0.4121 & 0.2400 & 0.2499 & 0.2185 & 0.1821 \\ 
                1.3636 & 0.3761 & 0.3265 & 0.2517 & 0.2448 & 0.1505 & 0.1610 \\ 
                1.0000 & 0.2749 & 0.2118 & 0.1736 & 0.1475 & 0.1204 & 0.0974 \\ 
                0.7500 & 0.2241 & 0.1572 & 0.1311 & 0.1045 & 0.0905 & 0.0683 \\ 
                0.5000 & 0.1768 & 0.1326 & 0.1020 & 0.0871 & 0.0641 & 0.0530 \\  
                0.2727 & 0.1547 & 0.0977 & 0.0788 & 0.0605 & 0.0453 & 0.0351 \\ 
                0.1429 & 0.1308 & 0.0740 & 0.0540 & 0.0449 & 0.0337 & 0.0248 \\ 
                   \hline
            \end{tabular}
            }
			\caption{$\langle a_i^2 \rangle / \langle D_t u_i^2 \rangle$ single cartesian component particle acceleration variance normalized by the fluid-tracer acceleration variance for various $\beta$ and $d/\eta$ values at $Re_{\lambda} = 32$.}
			\label{table:acce_variance}
		\end{center}
	\end{table}
	\begin{figure}
		\begin{center}	
		\includegraphics[width=0.9\columnwidth]{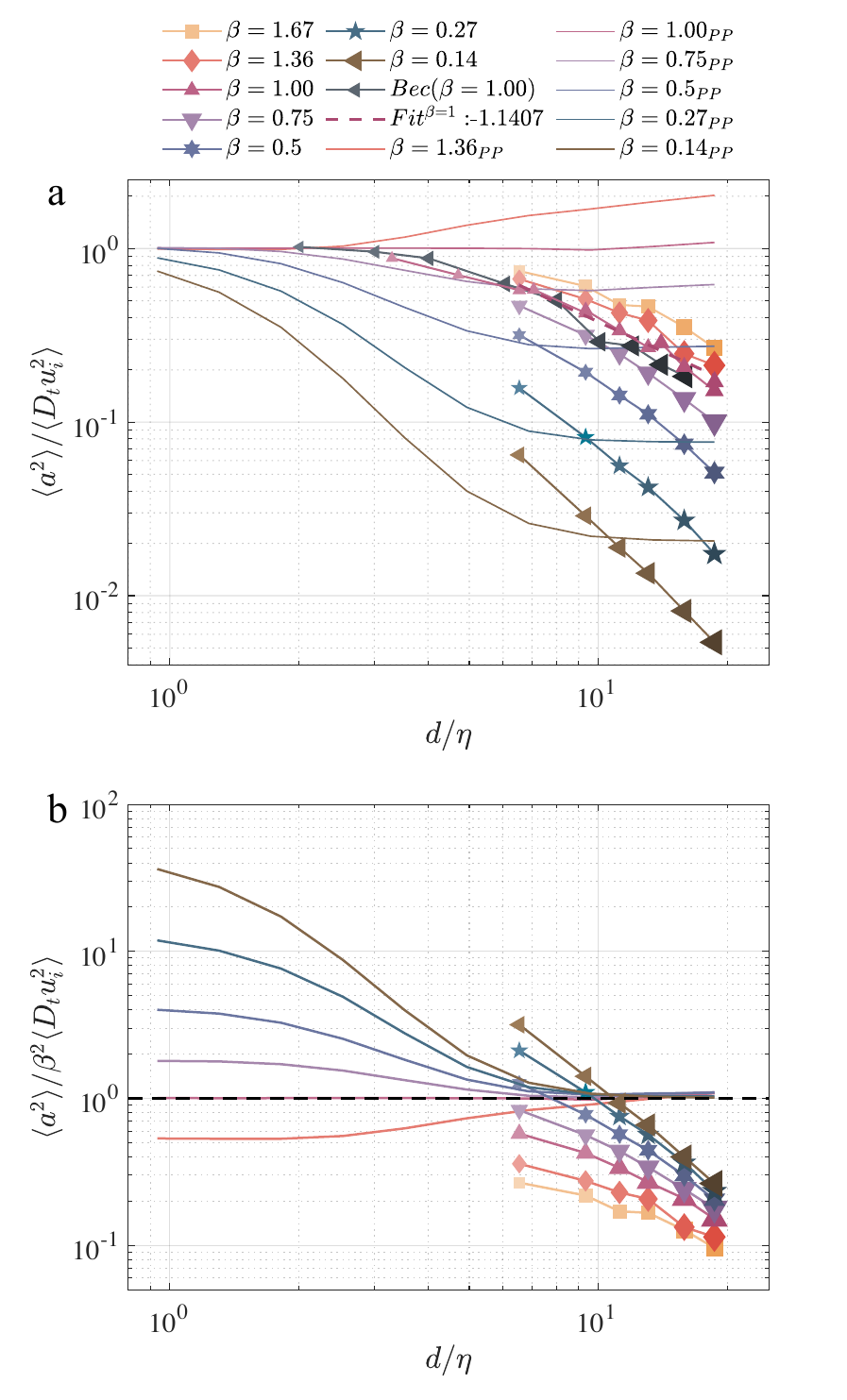}
			\caption{
   (a) Particle acceleration variance normalized by fluid-tracer acceleration variance as a function of  the particle diameter ($d$) in dissipative units $\eta$.
   Data for density ratios $\beta$ are displayed. The corresponding results for the point-point particle (PP) model are reported (solid line). Bec ($\beta=1$) indicates numerical measurements from \cite{homann_bec_2010}.
   (b) Same data with the normalization $ \beta^2 \langle (D_t u_i)^2 \rangle$ for the acceleration variance, this way the PP model tends asymptotically to the unit value. Note that the fully-resolved data with different $\beta$ do not overlap.
			}
			\label{fig:a2}			
		\end{center}
	\end{figure}
Figure \ref{fig:a2}(a) displays the trends for such a quantity (normalized by the fluid flow acceleration variance) with respect to the particle diameter (in dissipative scale units) and for particle sets with different density ratios ($\beta$). For the neutral case, $\beta=1$, the acceleration intensity progressively decreases from the unit value for growing particle diameters, this is due to the spatial filtering effect mentioned above. This feature has been already highlighted in experiments \cite{Brown2009prl,volk2011dynamics} and also reproduced in fully resolved simulations \cite{homann_bec_2010,jiang_wang_liu_sun_calzavarini_2022}. 
A one dimensional estimate based on the Kolmogorov 1941 (K41) turbulence theory suggests that  $\langle a_{d,i}^2 \rangle \sim\langle  \langle D_t u_i\rangle_V^2  \rangle\sim ((\delta_d u)^2/d)^2 \sim  d^{-2/3}$, where $\delta_d u$ stands for a typical increment of velocity over a scale $d$. It has been shown both  experimentally and numerically that $\langle a_d^2 \rangle \sim d^{\alpha}$ with the exponent $\alpha$ varying between $-4/3$ at low Reynolds (i.e. $Re_{\lambda} \leq 100$ as here) \cite{homann_bec_2010,jiang_wang_liu_sun_calzavarini_2022} and $-2/3$ for developed turbulence \cite{Brown2009prl,volk2011dynamics}.\\
The measurements for $\beta \neq 1$, which represent the main novelty of the present study, show that the particle acceleration grows with $\beta$. In order to contrast the density-ratio and the particle-size effect we also trace on Fig.\ref{fig:a2}(a) the results for point-like particles (solid lines). The point-particle (PP) model includes the added mass but does not account for the spatial filtering. It can be obtained from (\ref{eq:faxen}) by replacing the volume and surface averages by the point values and by setting $c(Re_d) =1 $ \cite{calzavarini2009acceleration}.
While in the limit of small particles the PP model departs weakly from the fluid acceleration value (as suggested by eq.~(\ref{eq:acc-small})), in the large size limit it converges to  $\langle a_d^2 \rangle \simeq \beta^2 \langle (D_t u_i)^2 \rangle $\cite{calzavarini2009acceleration}. This is better evidenced in Fig.\ref{fig:a2}(b), where the acceleration variances are normalized by $\beta^2 \langle (D_t u_i)^2 \rangle$ and all the PP  data converge to the same plateau. Conversely, the fully resolved simulation results tend to vanish in the asymptotic limit due to the progressively enhanced smoothing by filtering of turbulent fluctuations.\\
The effect of fluid-to-particle density ratio on the acceleration variance  $\langle a_{d}^2 \rangle$ can be examined by dividing it by the acceleration variance of the neutral particle of the corresponding size $\langle a_{d,\beta=1}^2 \rangle$. This is shown in Fig. \ref{fig:a2b} (a), one can notice that the data still depend on the values of $\beta$.
At this point is worth testing the hypothesis, 
\begin{equation}\label{eq:hyp1}
\langle a^2_d \rangle \simeq \beta^2 \langle a^2_{d,\beta=1} \rangle.
\end{equation}
This is done in Fig.\ref{fig:a2b}(b). Although the trend suggests that the curves might overlap for very large-particles, beyond the integral scale of turbulence, $L \simeq 24 \eta$, the lack of collapse of the curves can be interpreted as an evidence of the non-multiplicative effect of the added mass and spatial filtering on inertial-scale particles. 
We note that this multiplicative effect was instead true in the FC model studied in \cite{calzavarini2009acceleration}, see in particular their Fig. 1(a) where with a similar rescaling all data collapse for large particle diameters. The origin of this discrepancy with respect to the FC model remains to be understood. In particular it would be interesting to check if it is due to the non-linear drag. Indeed the Shiller-Naumann correction was not included in \cite{calzavarini2009acceleration}. Although it was considered, only for neutrally buoyant particles, in the FC model studied in \cite{calzavarini2012impact}  where it was found to have a negligible effect on the acceleration properties. 
	\begin{figure}
		\begin{center}	
              \includegraphics[width=0.9\columnwidth]{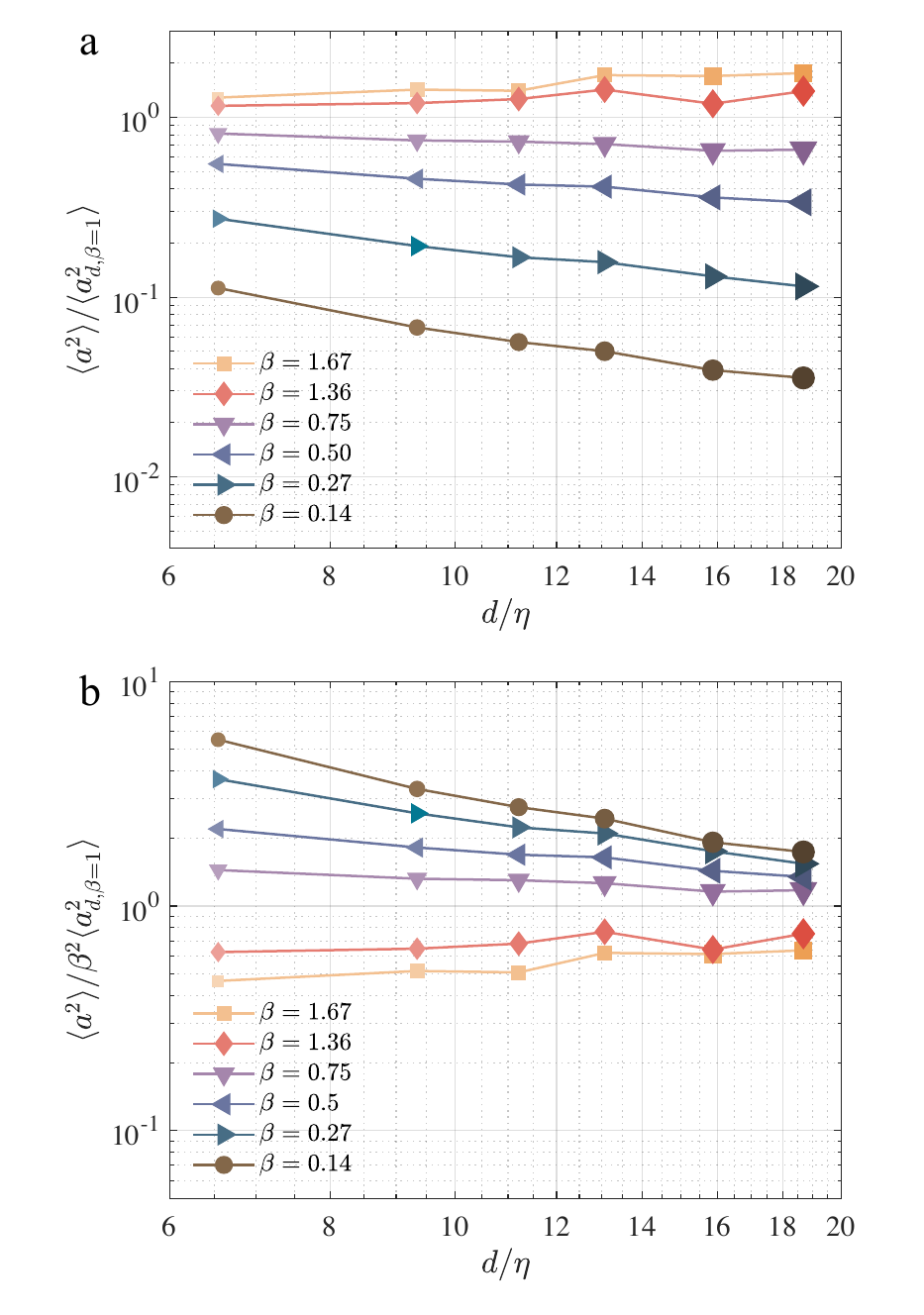}	

			\caption{
   (a) Particle acceleration variance normalized by the acceleration variance of the neutrally buoyant particle of the corresponding size $\langle a_{d,\beta=1}^2 \rangle$ as a function of  the particle diameter ($d$) in dissipative units $\eta$. Note that the curves as approximately horizontal for same $\beta$ particle families. 
   (b) Same data with the normalization $ \langle a_{d,\beta=1}^2 \rangle$ for the acceleration variance. Although curves get closer, they do not collapse on each other.
			}
			\label{fig:a2b}			
		\end{center}
	\end{figure}

	\begin{figure}
		\begin{center}	
	 \includegraphics[width=0.8\columnwidth]{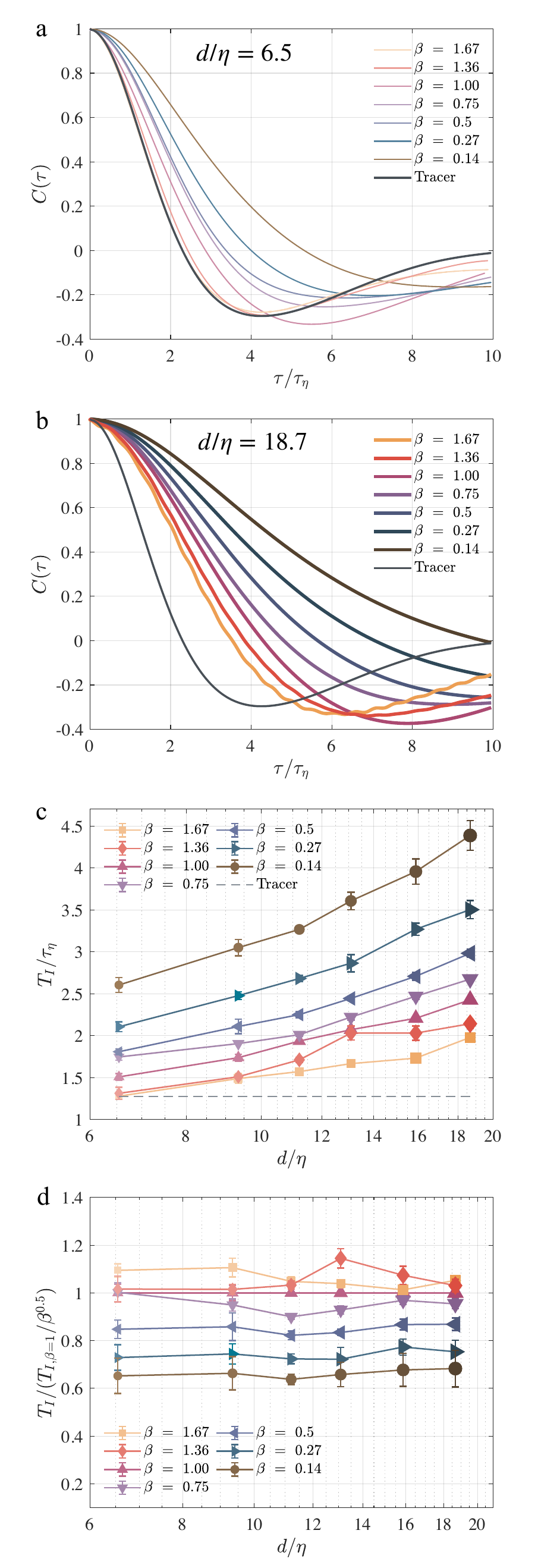}
	     
			\caption{
   Single component correlation function acceleration variance for particles diameter (a) $d/\eta = 6.5$ and (b) $d/\eta = 18.7$.
   (c) Integral correlation time of the particle acceleration versus the particle size, both expressed in dissipative units, for different $\beta$ families.
   (d)  Integral correlation time divided by $T_{I,\beta=1} /\sqrt{\beta}$ versus the particle size.}
			\label{fig:Ca}			
		\end{center}
	\end{figure}

\subsection{Acceleration temporal correlations}
The study of the temporal correlation functions of the acceleration allows to further reinforce the observations made for the instantaneous acceleration variance. 
We focus on the integral time, $T_I$, here defined as the integral of the autocorrelation function from time zero to the time it first reaches the null value (i.e. first zero-crossing time $T_0$):
	\begin{equation}
		T_{I}\equiv\int_{0}^{T_0}C(\tau)d\tau,~~ C(\tau)\equiv\frac{\langle a_i(t+\tau)a_i(t)\rangle}{\langle(a_i(t))^2\rangle}
		\label{eq:autoco}
	\end{equation}
 Figure \ref{fig:Ca}(a,b) shows the $C(\tau)$ correlations functions for the two small/large particle limiting case in this study, i.e. for (a) $d/\eta = 6.5$ and (b) $d/\eta = 18.7$.  
	\begin{table}
		\begin{center}
        \resizebox{\columnwidth}{!}{
            \begin{tabular}{|c|cccccc|}
            \hline
                \diagbox{$\beta$}{$d/\eta$} & 6.5419 & 9.3444 & 11.2119 & 13.0785 & 15.8774 & 18.6739 \\ \hline
                1.6667 & 1.2126	& 1.5015 & 1.6117 & 1.6425 & 1.7564 & 2.0205 \\
                1.3636 & 1.4170	& 1.5184 & 1.6928 & 1.9592 & 1.9589	& 2.1629 \\ 
                1.0000 & 1.6263	& 1.7011 & 1.9216 & 2.0704 & 2.2102	& 2.4419 \\ 
                0.7500 & 1.7408	& 1.8971 & 2.0315 & 2.2539 & 2.5249	& 2.6728 \\ 
                0.5000 & 1.8086	& 2.0344 & 2.2258 & 2.4176 & 2.6733	& 3.0411 \\  
                0.2727 & 2.1082	& 2.5053 & 2.6268 & 2.7740 & 3.2323	& 3.6138 \\ 
                0.1429 & 2.7127	& 3.1644 & 3.3539 & 3.5064 & 4.1050	& 4.5701 \\ 
                \hline
            \end{tabular}
            }
			\caption{$T_I/\eta$, integral correlation time of the particle acceleration normalized by the Kolmogorov time scale for various $\beta$ and $d/\eta$ values at $Re_{\lambda} = 32$.}
			\label{table:T_l}
		\end{center}
	\end{table}
The integral correlation times computed from these curves grow with the particle size and decreases with $\beta$, see  Fig. \ref{fig:Ca}(c) (all $T_I$ measurements are also reported in Table \ref{table:T_l}).  The first trend is usually rationalized in term of the coarse-graining hypothesis \cite{qureshi2007turbulent}. 
A particle of size $d$ is subjected to turbulence fluctuations of that scale, this corresponds to an eddy turnover time $\tau_d = d /\delta_d u \sim d^{2/3}$.
This prediction is only approximately true for neutrally buoyant particles.  In fact, similarly to the trends observed for the acceleration variance also in the case of the correlation time the measured scaling $\tau_d \sim d^{\gamma}$ has a senstive Reynolds dependence, it is observed that $\gamma \leq 2/3$ at small Taylor-Reynolds number \cite{homann_bec_2010,jiang_wang_liu_sun_calzavarini_2022}, while it is $2/3 \leq \gamma \leq 1$ at large Reynolds \cite{volk2011dynamics}.

The above argument can not be straightforwardly extended to non-neutral particles. 
A possible approximate adaptation is presented in the following. We indicate with $X_a = \frac{1}{2} \langle D_tu^2 \rangle^{1/2} \tau_a^2$ the length spanned by a fluid particle over the time ($\tau_a$)  during  which the acceleration is correlated (and so approximately constant). Such a length can be travelled by a finite sized inertial particle over a time $\tau_d = \sqrt{2 X_a / \langle a_d^2 \rangle^{1/2}}$. Now using the hypothesis (\ref{eq:hyp1})  one obtains 
\begin{equation}
\tau_d = \frac{\tau_{d, \beta=1}}{\sqrt{\beta}} \sim \frac{d^{2/3}}{\sqrt{\beta}}.
\end{equation}
This prediction is tested in Fig. \ref{fig:Ca}(d), we observe that while the size dependency of the correlation time for any particle seems to be properly normalized by the neutral case (i.e. the coarse graining hypothesis holds true), the density dependence is only approximately explained in terms of the above scaling. The collapse is  better for the case of light particles were the above argument is more fitting. 

\subsection{Acceleration's higher statistical moments}
\begin{figure}
	\begin{center}
	   \includegraphics[width=0.9\columnwidth]{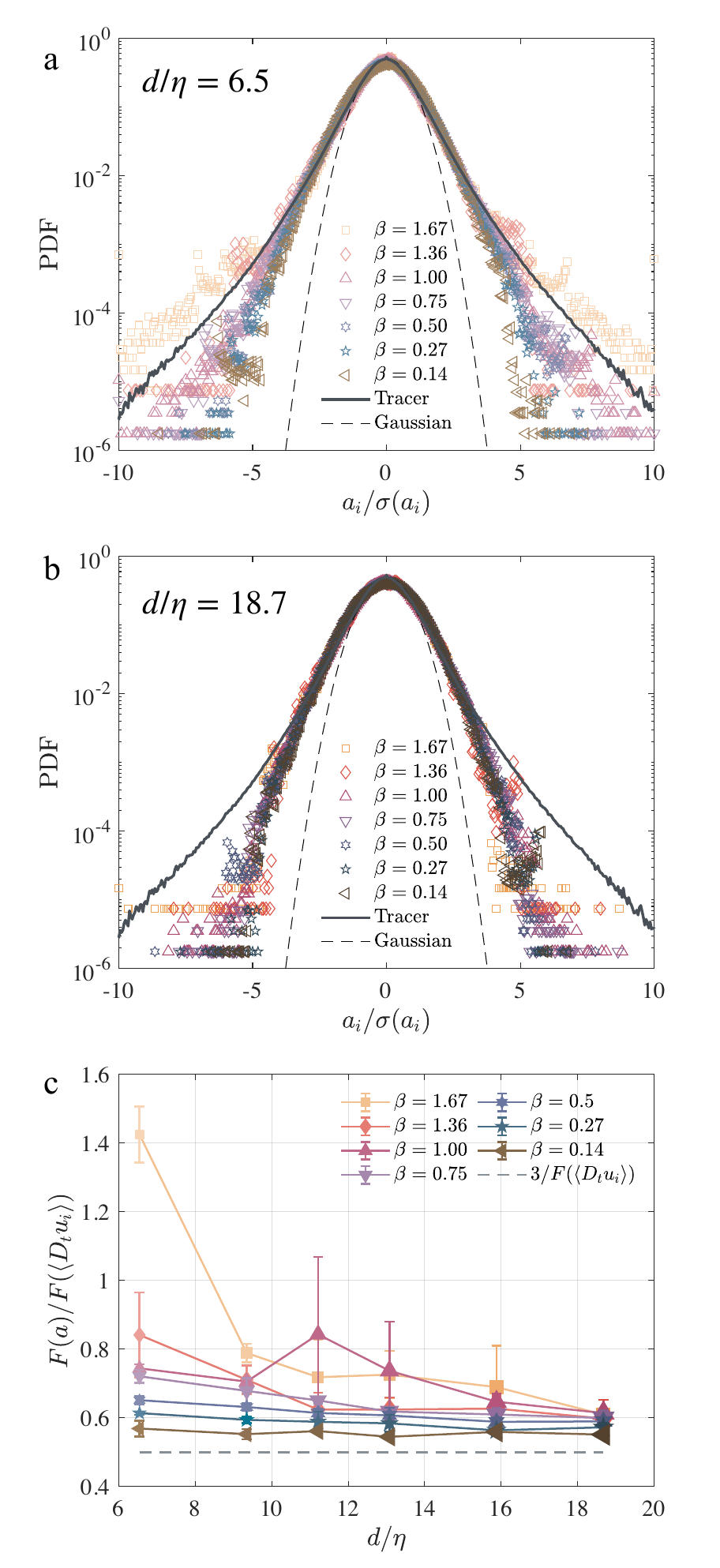}

			\caption{
PDF of particle acceleration normalized by their standard deviation for (a) $d/\eta = 6.5$ and (b) $d/\eta = 18.7$. \textcolor{black}{PDF of fluid tracers acceleration (bold solid line) and Gaussian distribution (dashed). (c) Flatness of the acceleration of particles versus their size. Data are normalized by the fluid acceleration flatness, $F(D_t u_i) \simeq 6$. The dashed line corresponds to the Gaussian value. Error bars are estimated from the semidifference of the flatness factors computed from two half-samples of the full dataset.}
			}	
			\label{fig:fa}			
		\end{center}
	\end{figure}

Last we consider statistical properties beyond the second order. This can be done by examining the trends in the shape of the probability density functions (PDF) of the acceleration normalized by its standard deviation. These functions are reported in Figure \ref{fig:fa}; panel (a) shows the curves for the smallest particles, while (b) for the largest particles explored in this study. It is clear the trend towards a Gaussianization of the accelerations when the particle size is increased, this feature was also predicted by the FC model simulations \cite{calzavarini2009acceleration}. A similar tendency is observed at increasing the particle mass density: PDF of heavy particles have shorter tails than neutral particles of the same size, on the contrary light particles tend to have extreme accelerations. This trend is evident for the smallest particles, and agrees with former results from PP model simulations \cite{VOLK20082084}, while it appears here negligible for the largest particles.\\ 
More robust conclusions can be drawn from the flatness of the acceleration, $F(a_d) = \langle a_d^4 \rangle / \langle a_d^2 \rangle^2$.
Theoretical considerations suggests that  $ \langle a_d^4 \rangle / \langle a_d^2 \rangle^2 \sim \langle (\delta_d u)^8  \rangle /  \langle (\delta_d u)^4  \rangle ^2 \sim d^{\zeta_8 - 2\zeta_4 < 0 }$, where  $\zeta_p$ indicates the scaling exponent of the velocity structure functions of order $p$, i.e.,  $\langle (\delta_d u)^p \rangle \sim d^{\zeta_p}$. The value of the exponent $F(a_d) \sim d^{\phi}$ can be estimated in various ways, it is $\phi \simeq -0.44$  from  Kolmogorov-Obukhov 1962 model \cite{kolmogorov1962} and  $\phi \simeq -0.56$ with She-L\'ev\^{e}que parametrization \cite{SheLeveque1994}.
Measuring a scaling behaviour for $F(a_d)$ is delicate as it requires large datasets. Furthermore, similar to the acceleration variance, large Reynolds numbers are needed in order to establish a wide inertial-range.  Although some experiments and simulations did not observe any scaling trend ($\phi=0$)  \cite{qureshi2007turbulent,xu2008motion, Brown2009prl,homann_bec_2010}, there are recently increasing evidences towards a reduction of the flatness with size,  \cite{volk2011dynamics} measured $\phi \simeq -0.5 \pm 0.1$ for inertial range neutrally buoyant particles, \cite{Prakash_2012} observed it for bubbles in turbulence. This trend is also supported by the FC model.
The reduction of flatness with particle size increase is confirmed by the present fully resolved simulations (see Figure \ref{fig:fa} panel (c)). 
Furthermore, we observe a weak but detectable flatness amplification with $\beta$.  This was previously observed only for dissipative scale particles both in experiments and simulations \cite{VOLK20082084}. Its physical interpretation lies in the phenomenon of preferential sampling: light particles explore preferentially vortex cores where the fluid centripetal acceleration is large, while heavy particles sample the outside of vortices which are calmer regions as far as acceleration is concerned, and this is reflected on their normalized accelerations PDF \cite{bec_biferale_boffetta_celani_cencini_lanotte_musacchio_toschi_2006,calzavarini2008dimensionality}.  This points to the fact that preferential sampling is a small but non vanishing effect for inertial-scale particles.

\section{Conclusions and Perspectives}
Non-interacting finite-sized massive spherical particles advected by a vigorous turbulent flow have been studied by means of fully-resolved numerical simulations. We examined the single- and two-time statistical properties of particle accelerations and explored their behaviours with respect to the particle size and to the particle mass density. 
In this conditions the inertial forces dominate over the dissipative ones.  This study confirms all the tendencies predicted by the coarse graining picture, namely the fact that both the acceleration variance and its flatness decrease with the particle size and the opposite for the acceleration correlation time. Given the limited extension of the inertial range at the present Reynolds number ($Re_{\lambda}=32$) it is difficult to estimate reliable scaling laws as a function of the particle diameter $d$. However, it seems that the observed behaviours deviate systematically from the expected scaling laws derived from the values of the velocity structure function exponents $\zeta_p$. This point is delicate and requires further attention. It might be  that the scaling laws derived from $\zeta_p$ only holds asymptotically in $Re_{\lambda}$, or it could be that systematic subleading deviations are present due to the influence of other hydrodynamics effects (such as the drag).\\
Looking at the particle density dependence, we have tried to assess the hypothesis  $a_d \sim \beta \langle D_t u_i \rangle_V$. 
Even if the particle acceleration increases with $\beta$, the observed trends can not be interpreted as the simple multiplicative combination of the two dominant terms: the spatial filtering of fluid accelerations and the $\beta$ fluid-to-particle density ratio. 
A similar mismatch is observed in the dependence of the correlation time with $\beta$, for which we provided a simple model. The study of the acceleration flatness indicates that light particles are more intermittent than heavy ones and this also when their size is large. This features suggest a role of preferential sampling of the flow by the particles. This interpretation shall be put under scrutiny in further studies, in fact in the context of inertial-scale particles the evidences of preferential sampling and the related preferential clustering are still not univocal \cite{FiabanePRE2012,uhlmann_chouippe_2017}.

In the future it will be interesting to perform simulations at larger Reynolds number and in larger domain size in order reduce the impact of finite inertial-range and finite-domain effects present in our simulations. It will also be interesting to develop new analysis techniques for the detection of preferential sampling by inertial-scale particles in turbulent flows. 
Furthermore, the statistical relevance for large particles in turbulence of forces such as history \cite{Govindarajan_history2023} or lift \cite{Mazzitelli2004}, remains to be clearly assessed.

These results may help in developing effective models for the dynamics of large particles in different context where particles are large with respect to the typical variation scale in the flow, such drifters and floaters in the ocean \cite{beronveraND2021} or rock crystals in magmatic chambers and primordial magma oceans \cite{PATOCKA2022117622}.

\section{Acknowledgement}
This work was supported by the National Natural Science Foundation of China under grant no. 11988102, New Cornerstone Science Foundation through the New Cornerstone Investigator Program and the XPLORER PRIZE.


\end{document}